\documentclass[10pt,english, aps,superscriptaddress,showpacs,floatfix,prl,lengthcheck,floatfix,nofootinbib,twocolumn]{revtex4-1}

\usepackage[T1]{fontenc}
\usepackage[latin9]{inputenc}
\usepackage{color}
\usepackage{amstext}
\usepackage{graphicx}

\makeatletter
\usepackage[latin9]{inputenc} %%% acentos no bibtex resolve a questao

\usepackage{braket}

\usepackage[T1]{fontenc}
\usepackage{color}
\usepackage{babel}
\usepackage{verbatim}
\usepackage{amssymb,amsfonts,amsmath}
\usepackage{graphicx}
\usepackage[pdfusetitle,colorlinks,linkcolor=blue,citecolor=blue,urlcolor=blue]{hyperref}
\usepackage{microtype}

\makeatother

\usepackage{babel}
\begin{document}
\title{High-Harmonic Generation with a twist: all-optical characterization
of magic-angle twisted bilayer graphene.}
\author{Eduardo B. Molinero}
\email{ebmolinero@gmail.com}

\affiliation{\emph{Instituto de Ciencia de Materiales de Madrid (ICMM), Consejo
Superior de Investigaciones Científicas (CSIC), Sor Juana Inés de
la Cruz 3, 28049 Madrid, Spain}}
\author{Anushree Datta}
\affiliation{\emph{Instituto de Ciencia de Materiales de Madrid (ICMM), Consejo
Superior de Investigaciones Científicas (CSIC), Sor Juana Inés de
la Cruz 3, 28049 Madrid, Spain}}
\author{M. J. Calderón}
\affiliation{\emph{Instituto de Ciencia de Materiales de Madrid (ICMM), Consejo
Superior de Investigaciones Científicas (CSIC), Sor Juana Inés de
la Cruz 3, 28049 Madrid, Spain}}
\author{E. Bascones}
\affiliation{\emph{Instituto de Ciencia de Materiales de Madrid (ICMM), Consejo
Superior de Investigaciones Científicas (CSIC), Sor Juana Inés de
la Cruz 3, 28049 Madrid, Spain}}
\author{Rui E. F. Silva}
\email{ruiefdasilva@gmail.com}

\email{rui.silva@csic.com}

\affiliation{\emph{Instituto de Ciencia de Materiales de Madrid (ICMM), Consejo
Superior de Investigaciones Científicas (CSIC), Sor Juana Inés de
la Cruz 3, 28049 Madrid, Spain}}
\affiliation{\emph{Max Born Institute, Max-Born-Strasse 2A, D-12489 Berlin, Germany}}
\begin{abstract}
If we stack up two layers of graphene while changing their respective
orientation by some twisting angle, we end up with a system that has
striking differences when compared to single-layer graphene. For a
very specific value of this twist angle, known as \emph{magic angle},
twisted bilayer graphene displays a unique phase diagram that cannot
be found in other systems. Recently, high harmonic generation spectroscopy
has been successfully applied to elucidate the electronic properties
of quantum materials. The purpose of the present work is to exploit
the nonlinear optical response of magic-angle twisted bilayer graphene
to unveil its electronic properties. We show that the band structure
of magic-angle twisted bilayer graphene is imprinted onto its high-harmonic
spectrum. Specifically, we observe a drastic decrease of harmonic
signal as we approach the magic angle. Our results show that high
harmonic generation can be used as a spectroscopy tool for measuring
the twist angle and also the electronic properties of twisted bilayer
graphene, paving the way for an all-optical characterization of moiré
materials.
\end{abstract}
\maketitle

\section*{Introduction}

The discovery of magic-angle twisted bilayer graphene (MATBG) \citep{Cao2018_matbg_I,Cao2018_matbg_II}
has sparked a huge increase of attraction into moiré quantum materials
\citep{Andrei2021} in the recent years. One of the key reasons for
that increase is that MATBG shows a unique plethora of exotic phenomena
\citep{Andrei2020}, which includes correlated insulators \citep{Cao2018_matbg_I,Lu2019},
unconventional superconductivity \citep{Cao2018_matbg_II,Yankowitz2019,Lu2019},
interacting topological phases \citep{Nuckolls2020_chern_matbg,Choi2021_topo_matbg,XieNature2021},
ferromagnetism \citep{aaron2019_ferro_matbg} and also strange metal
behavior \citep{cao2020-strange-metal,JaouiNatPhys2022}. Such phase
diagram has some resemblance with the phenomenology present in cuprates
\citep{Keimer2015} and iron pnictides \citep{Fernandes2022} superconductors.
For instance, both systems share the presence of superconducting domes
surrounded by correlated insulating states as doping is varied. Moiré
materials have opened completely new avenues in research thanks to
the unprecedented in-situ tunability of their electronic properties
through electronic gates, strain, magnetic field, pressure or substrate
alignement among others. This has for example allowed observing many
different correlated insulating states in a single MATBG sample, including
integer and fractional Chern insulators as well as trivial insulators
with or without charge density modulations \citep{XieNature2021}.

At the heart of the correlated phases of MATBG it is the existence
of two flat bands \citep{Bistritzer2011,Lisi2021,lopesdossantos2007}
at the so-called magic angle $\theta_{\text{mag}}\approx1.1^{\circ}$.
As the bandwidth is reduced, the role played by the interactions increases
\citep{Choi2019,DattaArXiv2023} and rich and complex physics arises.
The nature of these flat bands is far from trivial, as they are formed
by the hybridization of the Dirac cones of each layer of graphene
\citep{Cao2018_matbg_I,san-jose2012}. A proper characterization of
the low energy bandstructure of MATBG is key to unveil the origin
of the correlated states. On the other hand, studying the electronic
properties of these flat bands requires of new spectroscopic tools
as the large moiré unit cell and the two dimensionality of MATBG complicate
the applicability of many experimental techniques of common use in
other strongly correlated materials.

\begin{figure}
\begin{centering}
\includegraphics[width=0.8\columnwidth]{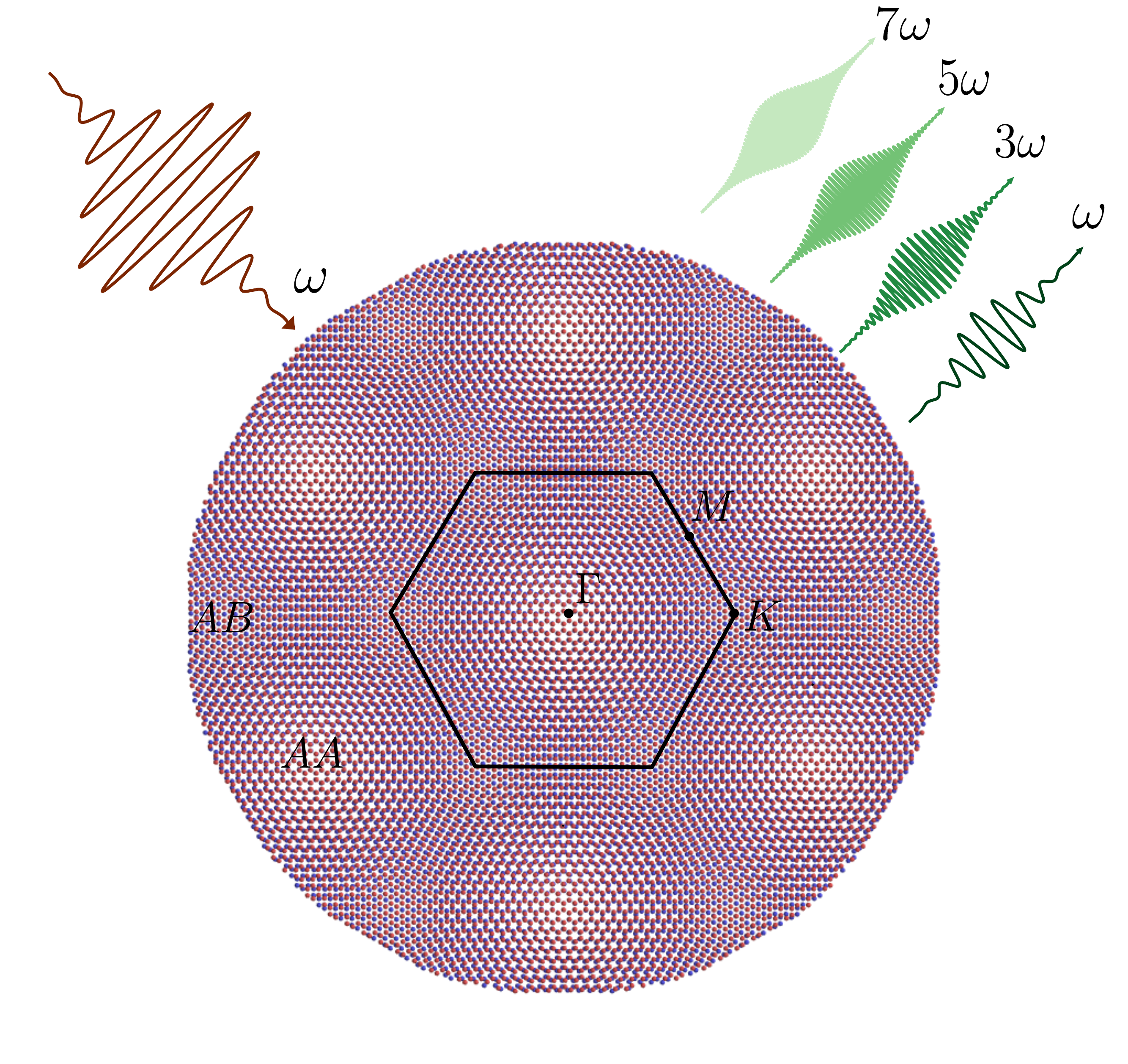}
\par\end{centering}
\caption{\label{fig:1_laser}Schematic figure of the moiré superlattice alongside
with the laser directions. Note that the $\Gamma-M$ direction connects
$AA$ to $AA$ zones while $\Gamma-K$ direction connects $AA$ to
$AB/BA$ zones. We note that the moiré Brillouin zone is not shown
at scale.}
\end{figure}

In parallel, the study of the interaction between ultrashort laser
pulses and condensed matter systems has also experienced a huge growth
in relevance \citep{kruchinin2018_colloquium}. Such short pulses
can drive the system's electrons into a non-equilibrium excited state
within times below their own cycle. As a consequence of this excitation,
the electrons emit coherent radiation in multiples (up to the hundreds)
of the laser frequency, providing us a tool to unveil their dynamics.
Even though this high-harmonic generation (HHG) had its origins in
atomic and molecular physics \citep{krausz2009}, it has been recently,
and successfully, applied to condensed matter systems as a spectroscopy
tool \citep{Ghimire2018,Goulielmakis2022}. For instance, it has allowed
for the observation of ultrafast electron-hole dynamics \citep{Vampa2015},
insulator-to-metal transitions \citep{Silva2018_mott,bionta2021},
topological transitions \citep{Silva2019_topological,chacon202},
and light-driven band structure \citep{UzanNarovlansky2022}.

\begin{figure*}
\includegraphics[width=2\columnwidth]{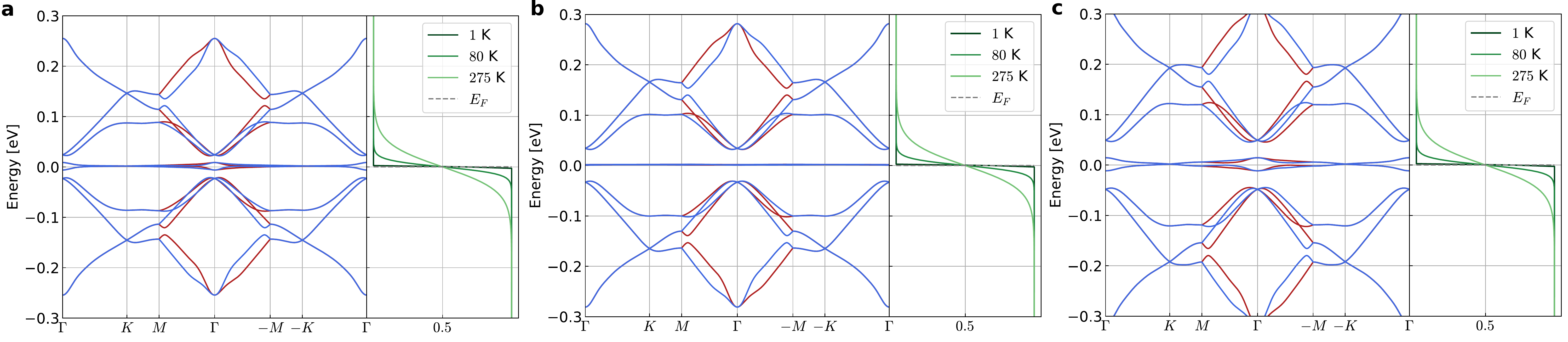}

\caption{\label{fig:2_bands} Band structure of TBG for 3 distinct twist angles
alongside its Fermi-Dirac distribution. \textbf{a} corresponds to
$\theta=1.05^{\circ}$, \textbf{b} to $\theta=1.12^{\circ}$ and \textbf{c}
to $\theta=1.22^{\circ}$, showing the appearance of flat bands at
the so-called magic angle $\theta=1.12^{\circ}$. Red color indicates
the band structure associated with the $K$ valley, while blue lines
depicts its $K'$ counterpart (see Methods for a discussion on the
model).}
\end{figure*}

Why HHG can be effectively used as a spectroscopy tool, can be understood
thanks to its underlying physical mechanisms. The HHG spectrum in
solids comes from two distinct contributions, namely intraband and
interband currents. Intraband harmonics are generated due to the accelerated
motion of the electron (hole) in the conduction (valence) band after
a multiphoton/tunneling excitation of the electron from the valence
band. Afterwards, the recombination of such pair gives rise to the
interband harmonics. While intraband contributions are generally associated
with low order harmonics \citep{Jrgens2020}, higher-order ones are
caused by the latter contributions \citep{vampa2014}. In the presence
of a strong low frequency laser field, the charge carriers are drifted
through the whole Brillouin zone, exploring sections, and hence physics,
that are not commonly seen by other optical tests, such as second
harmonic generation \citep{boyd2020nonlinear}.

In the present work, we aim to bridge the gap between these two distinct
fields by studying HHG in magic-angle twisted bilayer graphene. Remarkably,
in spite of extensive work in twisted bilayer graphene \citep{bernevig2021_TBG_I,bernevig2021_TBG_II,bernevig2021_TBG_III,bernevig2021_TBG_IV,bernevig2021_TBG_V,bernevig2021_TBG_VI},
its ultrafast dynamics have been hardly explored, with very few exceptions
\citep{ikeda2020_hhg_tbg,mrudul2021_hhg_tbg,du2021_hhg_tbg,Yang2020}.
However and to the knowledge of the authors, no previous work has
dealt with highly nonlinear optical response in MATBG. Here we show
that the emitted spectrum reveals the formation of the flat bands
at the magic angle.

\section*{results}

As it has been discussed earlier, the HHG of solids is extremely dependent
on the electronic band structure of the system \citep{tancognedejean2017,silva2019_wannier},
up to the point that one can reconstruct the band structure from its
high-harmonic spectrum \citep{vampa2015-band-recons}. Therefore,
the harmonic spectra of twisted bilayer graphene must inevitably reflect
the existence (or absence) of the flat bands.

\begin{figure}
\includegraphics[width=1\columnwidth]{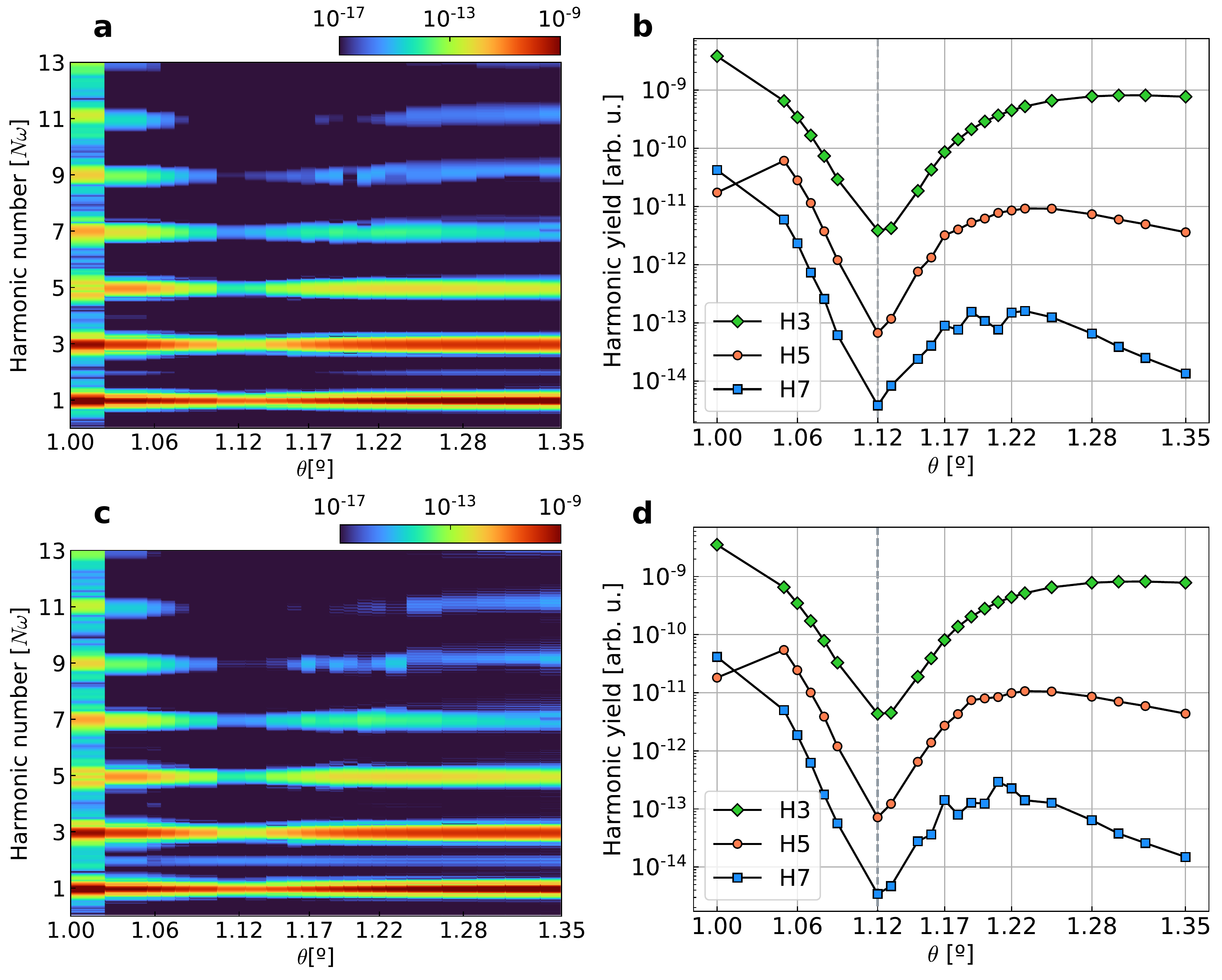}

\caption{\label{fig:3_harm} \textbf{a} and \textbf{c}, High harmonic spectrum
for a set of twist angles. \textbf{b} and \textbf{d}, Yield for the
third (H3), fifth (H5) and seventh (H7) harmonic. \textbf{a} and \textbf{c}
results are obtained for a laser in the $\Gamma-M$ direction, while
\textbf{c} and \textbf{d} are obtained for the same laser but in the
$\Gamma-K$ direction. All calculations have been done for a temperature
of $80$ K. One can see the clear depression of the harmonic signal,
for all subfigures, near the magic angle.}
\end{figure}

The purpose of this work is to show that one can characterize the
magic angle in an all-optical way, offering an alternative to other
characterization techniques, such as scanning tunneling microscopy
\citep{Lisi2021}. We show that when the twist angle of the TBG gets
near $\theta_{\text{mag}}$, the intensity of the high harmonics (from
the third upwards) is reduced by several orders of magnitude. Moreover,
it is shown that the effect is robust against important variations
of the laser field and to temperature changes, providing a more robust
probe of magic angle over the existing ones.To check this hypothesis,
we performed numerical simulations (see Methods for more details on
the model and the time propagation scheme). To test the effect of
the flat bands on the emission spectra, we will scan a set of twist
angles, near $\theta_{\text{mag}}$, ranging from $1.00^{\circ}$
to $1.35^{\circ}$. In Fig. \ref{fig:2_bands}, we show the band structure
for 3 different angles: the magic angle itself ($\theta=1.12^{\circ}$
in our model), one angle above ($\theta=1.22^{\circ}$) and one below
it ($\theta=1.05^{\circ}$). One can notice how the bands lose their
flatness when we move away from $\theta_{\text{mag}}$.

Figure \ref{fig:3_harm}(a) shows the harmonic spectrum of the TBG
for a laser in the $\Gamma-M$ direction (see Fig. \ref{fig:1_laser})
as a function of the twist angle. A main feature emerge: a strong
suppression of the intensity of the odd non-linear harmonics for twist
angles near $\theta_{{\rm mag}}$. In order to obtain a clearer grasp
of the magnitude of this decrease, Fig. \ref{fig:3_harm}(b) presents
the harmonic yield from the third up to the seventh harmonic. In spite
of the small angle differences, the intensities of each of the three
harmonics differ by several orders of magnitude. This disparity would
allow for all-optical characterization of the magic-angle twisted
bilayer graphene, even for twist angles close to each other; see for
instance, the difference between $1.06^{\circ}$ versus $1.12^{\circ}$.

Even though the process here described is rather complex, the physical
intuition behind it, is not. At a semiclassical level, the intraband
current can be expressed as \citep{vampa2014}
\begin{equation}
\mathbf{j}_{\text{intra}}(t)\propto\sum_{\mathbf{k},n}\nabla_{\mathbf{k}}\varepsilon_{n}(\mathbf{k}(t)),
\end{equation}
where $\varepsilon_{n}(\mathbf{k}(t))$ is the energy of the $n$-th
band and the crystal momentum is given by $\mathbf{k}(t)=\mathbf{k}+e\mathbf{A}(t)$.
Therefore, the most relevant contribution to the emission spectra
is the \emph{curvature} of the bands. Hence, when the band starts
to flatten, the harmonic intensity will be reduced. While this semiclassical
explanation cannot account for all the subtle, yet interesting, physics
of the system, it provides an understandable picture of this process.

Consequently, this phenomenon must be robust under important variations
of the laser because it is only based on the existence of the flat
bands. This intuition is confirmed by our numerical simulations. Figure
\ref{fig:3_harm}(c,d) shows the same quantities as in \ref{fig:2_bands}(a,b)
but for a laser in the $\Gamma-K$ direction, see Fig. \ref{fig:1_laser}.\textcolor{red}{{}
}Independently of the laser direction, the same physics appears; when
the twist angle gets close to $\theta_{\text{mag}}$, a depression
of several orders of magnitude appears in the emission spectrum.

As seen above, the suppression of the harmonic intensity close to
the magic angle crucially depends on the narrowing of the flat bands
but, what happens when other bands are involved on the dynamics? For
instance, when the upper conduction bands have some population due
to an increase in the temperature, see Fig. \ref{fig:2_bands}. Intuitively,
one would expect that when upper bands get populated, the current
will have relevant intraband contributions and hence, the decrease
on the emission will be smoothed out, as the harmonic signal will
not be exclusively originated from the flat bands.

\begin{figure}
\includegraphics[width=1\columnwidth]{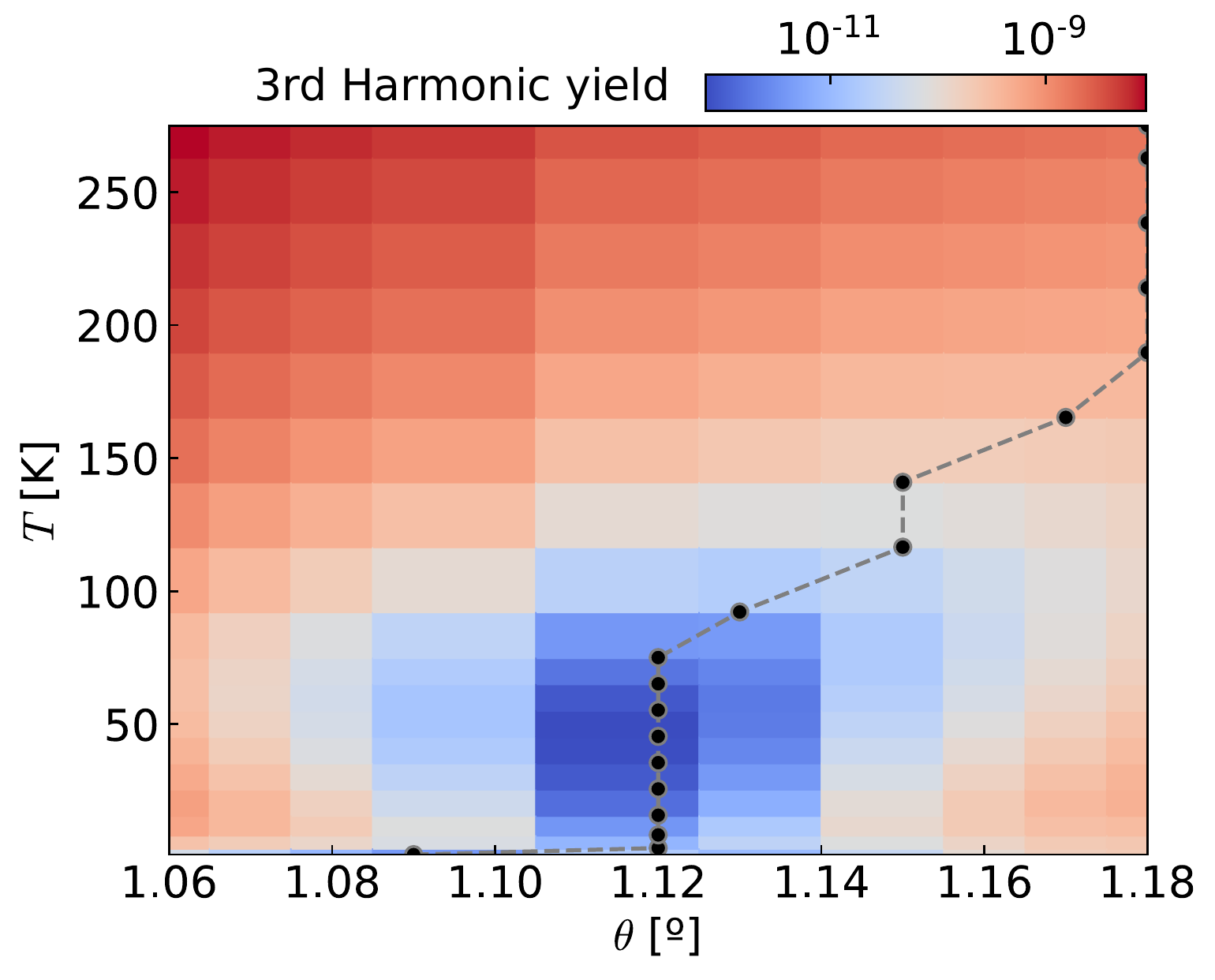}

\caption{\label{fig:4_temp}Yield of the third harmonic in terms of temperature.
Black circles depicts the minimum of the yield for each twist angle.}
\end{figure}

To check this hypothesis, we performed a wide temperature scan for
a set of angles near $\theta_{\text{mag}}$. Fig. \ref{fig:4_temp}
shows the yield for the 3rd harmonic in terms of the twist angle and
the temperature. For temperatures below $100$ K, the depression on
the intensity is heavily located around the magic angle. As seen from
the location of black circles which signify the minimum of the 3rd
harmonic, the minimum is located at the magic angle ($1.12^{\circ}$).
However, this depression near $\theta_{\text{mag}}$ disappears when
the temperature is enough to populate the upper bands, and the minimum
moves away from the magic-angle. This effect is still visible for
temperatures that are well above the ones at which the correlated
insulators and superconductivity have been observed in MATBG.

For low frequency laser pulses, the most important contributions to
the current, and therefore to the harmonic signal, come from bands
near the Fermi energy, in the case of TBG, the flat bands. In particular,
by choosing laser frequencies that are smaller than the typical bandwidth
of the flat bands in TBG (here $\omega=1\text{ meV}$), spurious contributions
from higher conductions bands can be avoided. Such small laser frequencies
are crucial for our study and to detect experimentally the drop in
the harmonic efficency due to the flattening of the bands. Laser frequencies
and intensities similar to the ones considered here have been used
in high harmonic generation measurements on a graphene monolayer \citep{Hafez2018},
making the experiments proposed experimentally feasible.

\section*{discussion}

Our results have shown that the flatness of the bands of MATBG has
a strong impact on its high harmonic spectrum. Their narrowing close
to the magic angle produces a decrease of several orders of magnitude
in the emission intensity. Such discrepancy in the optical response
can be used in a fully optical study of MATBG. This effect is crucially
based \emph{only} on the existence of the flat bands and it holds
for a wide range of laser parameters, even showing no anisotropy in
the laser direction. In addition, we have confirmed that the suppression
is maintained within a relatively wide range of temperatures up until
100 K, far away from the ultralow temperature regime used in most
MATBG experiments. Nevertheless, we hope that our work can not only
be useful as a spectroscopy tool but also, shed light onto the mechanisms
behind electronic ultrafast dynamics in twisted materials.

\section*{\label{sec:Methods}Methods}

To model TBG we start from the continuum model \citep{lopesdossantos2007,Bistritzer2011,san-jose2012}
taking a graphene Fermi velocity of $v_{F}=2.34a$ eV, where $a$
is the lattice constant of graphene, and a ratio between the interlayer
tunneling at AA and AB regions of $w_{0}/w_{1}=0.78$ \citep{Nam2017-koshino}.
With these parameters the magic angle is found at $\theta\approx1.12^{\circ}$.
In the continuum model the two valleys of graphene are assumed to
be uncoupled. The spectrum of TBG equals the sum of the contributions
of each valley. In each of the valleys the model satisfies the symmetries:
$\mathcal{C}_{3}$, $\mathcal{C}_{2}\mathcal{T}$ (the combined operation
involving time reversal symmetry $\mathcal{T}$ and $\mathcal{C}_{2}$),
and a valley preserving mirror symmetry $M$. Here $\mathcal{C}_{3}$
and $\mathcal{C}_{2}$, involve respectively $120^{\circ}$ and $180^{\circ}$
rotations with respect to an axis perpendicular to the TBG and $M$
is a layer exchanging two fold rotation around an axis parallel to
the TBG. We then adapt the Wannier function model of TBG with 8 orbitals
per valley \citep{carr2019_8bands,calderon2020} to approximate the
bandstructure model obtained with the continuum model. The resulting
tight binding model includes hopping up to a radius of 10 moiré lattice
constant. This Wannier model satisfies the symmetries of the continuum
model in each valley. The TBG is assumed to be undoped for all the
calculations by setting the chemical potential for each temperature.
Since we are interested in the optical response of the bilayer, we
need to assume some form for the position operator $\mathbf{r}$,
and we restrain ourselves to the diagonal approximation \citep{ibanez2022_tbapprox,silva2019_wannier}.

Inversion and time reversal symmetry are satisfied only when the two
valleys are included. The spectrum of the even harmonics is finite
for each valley but it vanishes when adding the contribution of both
valleys.

The time-dependent calculations were performed by solving the semiconductor
Bloch equations (SBE), using Maximally Localized Wannier functions
\citep{Marzari2012} that are well suited for the model describe above,
following the formalism presented in \citep{silva2019_wannier}. During
this work, we have considered the interaction of the twisted bilayer
graphene with a relatively weak $(E_{0}=1\cdot10^{5}\text{ V/m})$
pulse in the microwave regime $(\omega=1\text{ meV})$ and a gaussian
envelope with a full width at half-maximum of $23.5$ ps. This optical
regime was chosen so to properly isolate the flat bands, i.e., to
avoid transitions between them and the upper conduction bands. Otherwise,
such processes would give relevant interband contributions to the
spectrum. Nevertheless, the physical behavior is robust under variations
of both the amplitude and the frequency of the field. We note that,
even though the field amplitude may seem weak, for the case of MATBG
it will be relatively strong due to the reduced bandwidth of the system.
The reduced density matrix was propagated in the Wannier gauge using
a $100\times100$ Monkhorst--Pack grid and a time step of $2.41$
fs. We checked that the dynamics were properly converged for all numerical
parameters. The dephasing term \citep{vampa2014,silva2019_wannier}
was set to $T_{2}=1000$ fs $\approx\Omega/4$, where $\Omega$ is
the laser period. However, we checked that the same results were obtained
if the value of $T_{2}$ was varied.

\section*{Acknowledgments}

E. B. M. and R. E. F. S. acknowledge support from the fellowship LCF/BQ/PR21/11840008
from \textquotedblleft La Caixa\textquotedblright{} Foundation (ID
100010434). This research was supported by Grant PID2021-122769NB-I00
funded by MCIN/AEI/10.13039/501100011033.

\bibliographystyle{apsrev4-1}
\bibliography{hhg_tblg}

\end{document}